\documentclass[preprint,12pt,authoryear]{elsarticle}

\usepackage{amssymb}
\usepackage{amsthm}
\usepackage{amsmath}

\usepackage[left=2.5cm,right=2.5cm,top=2.5cm,bottom=2cm]{geometry}
\usepackage[overload]{empheq}
\usepackage[colorlinks,allcolors=blue]{hyperref}
\usepackage{graphicx}
\usepackage{adjustbox}
\usepackage{cleveref}
\usepackage{siunitx}
\usepackage{gensymb}
\usepackage{appendix}
\usepackage{subcaption}
\DeclareRobustCommand{\VEC}[1]{\boldsymbol{#1}}
\DeclareRobustCommand{\MAT}[1]{\boldsymbol{\mathbb{#1}}}
\newcommand{\strain}{\varepsilon}
\newcommand{\strainb}{\VEC{\varepsilon}}
\newcommand{\plb}{\VEC{p}}
\newcommand{\stress}{\sigma}
\newcommand{\stressb}{\VEC{\stress}}
\newcommand{\plforb}{\VEC{X}}
\newcommand{\plform}{X_m}
\newcommand{\idtwo}{\mathbf{I}}     
\newcommand{\tr}{\text{Tr}\hspace{0.5mm}}
\newtheorem*{remark}{Remark}

\usepackage{marginnote}

\begin{document}

\begin{frontmatter}

\title{Hyperelastic nature of the Hoek--Brown criterion}

\author[1]{I. Fontana}
\ead{ifontana@arizona.edu}
\author[2]{G. Bacquaert\corref{cor}}
\ead{goustan.bacquaert@edf.fr}
\author[3]{D. A. Di Pietro}
\author[2]{K. Kazymyrenko}
\cortext[cor]{Corresponding author} 

\affiliation[1]{organization={University of Arizona, Department of Mathematics},city={Tucson}, country={USA}}
\affiliation[2]{organization={Électricité de France, R\&D Division},city={Palaiseau}, country={France}}
\affiliation[3]{organization={Université de Montpellier, CNRS, Institut Montpelliérain Alexander Grothendieck, UMR 5149},city={Montpellier},country={France}}

\date{July 6, 2025}

\begin{abstract}
We propose a nonlinear elasto-plastic model, for which a specific class of hyperbolic elasticity arises as a straight consequence of the yield criterion invariance on the plasticity level. We superimpose this nonlinear elastic (or hyperelastic) behavior with plasticity obeying the associated flow rule. Interestingly, we find that a linear yield criterion on the thermodynamical force associated with plasticity results in a quadratic yield criterion in the stress space. This suggests a specific hyperelastic connection between Mohr--Coulomb and Hoek--Brown (or alternatively between Drucker--Prager and Pan--Hudson) yield criteria. We compare the elasto-plastic responses of standard tests for the Drucker--Prager yield criterion using either linear or the suggested hyperbolic elasticity. Notably, the nonlinear case stands out due to dilatancy saturation observed during cyclic loading in the triaxial compression test. We conclude this study with structural finite element simulations that clearly demonstrate the numerical applicability of the proposed model.
\end{abstract}

\end{frontmatter}

\section{Introduction}

One of the key points in the accurate description of rocks is the precise identification of their elasticity domain (or yield surface/criterion). Due to the softening behavior, not only rocks but also all analogous materials like concrete, clay, soil, or sand are commonly classified by their resistance to various mixed-mode loading \citep{Kupfer1973,Roscoe1970}. This approach is closely related to the basic safety rules in industrial/geotechnical applications, where it is often considered that geomaterials could exhibit unstable failure once the critical loading is reached \citep{Hoek1982}. \\

Throughout the last century, specific testing machines and corresponding measurement protocols have been established by national and international committees in order to harmonize and standardize the experimental characterization of the above-cited brittle materials. To mention only a few, the Brazilian tensile test \citep{Carneiro1943}, 
the \oe dometric, uni- and tri-axial compression tests are all well-documented experiments that are routinely executed to catalog material strength by spotting just some points of their multidimensional yield surfaces. This reduced ``single point'' vision of material resistance has been increasingly scrutinized in recent times, and multiple evolutions have been adopted. For instance, constant improvements in finite element software enable new kinds of modeling, with loadings going beyond the elasticity domain to explore a subtler post-peak behavior, most notably with the emergence of shear bands governed by the plastic flow. In the corresponding mechanical tests, the response of the material subjected to a set of pre-established loadings is analyzed during both the elastic and softening phases, enabling the full model parameter fitting. While the complexity of post-peak description could be reached through various theoretical formalisms, most of them still rely on the initial yield surface definition, and the question of this elasticity domain shape remains the cornerstone of any nonlinear model identification. Even if considerable progress has been achieved in recent decades, this precise identification of the yield surface remains a rather challenging task \citep{Lee2004}. \\

A common feature of geomaterials is their strong resistance to compressive loading. For some large-scale structures, like hydraulic dams or underground tunnel excavations, the construction material is naturally submitted to high levels of compression. For others, like nuclear confinement buildings or bridges, civil engineering concrete parts are preloaded to reach artificially an initial compression state by supplementary constraint of tension reinforcing steel tendons. In both cases, the property of higher compressive resistance is exploited on the industrial level with the aim of increasing global structure robustness. \\

According to the physical origin of geomaterials, a large variety of criteria defining the elasticity domain are employed in order to model their mechanical behavior. The simplest surface, admitting infinite resistance in hydrostatic compression, is the linear cone-shaped one. It was first introduced more than a century ago \citep{ Mohr1900} as the combination of Coulomb's friction hypothesis \citep{Coulomb1776} with Galileo--Rankine's tension cut-off principle. The initial principal stress description, which is commonly called the Mohr--Coulomb shape, was later generalized in the work of Drucker and Prager to a smoother deviator-trace relation \citep{Drucker1952}. The straight relation between shear and compressive loadings, which is the main signature of these linear criteria, has allowed the development of various constitutive relations based on the same simplified dependence \citep{Alejano2012, Labuz2012}. At the same time, early research conducted in the 1950s that aimed to apply the minimal concept of normal plastic flow to materials that were supposed to fail according to Mohr--Coulomb-like criteria was mainly unsuccessful. The primary criticism of these theories was their tendency to cause excessive dilatancy due to the inherent constant flow angle related exclusively to the friction coefficient. One way to overcome this difficulty is to enhance the model by incorporating a flow rule that is independent of the shape of the yield surface (see e.g. \citep{Vermeer1984}). \\

In the middle of the last century, with numerous large infrastructural projects ongoing, more complex criteria emerged as further experimental data became available. For wider ranges of loadings, the friction-type shear dependency seemed to be deflecting from the linear curve. Logically, a quadratic relation was to be explored first. Back in 1924, assuming the hypothesis of crack propagation via rapid growth of randomly distributed micro-flaws, Griffith had already obtained a theoretical justification of the parabolic yield shape \citep{Griffith1924}. Inspired by Griffith's model, Fairhurst proposed its empirical extension validated on the tensile Brazilian test \citep{Fairhurst1964}. In this spirit, in 1980, Evert Hoek and Edwin T. Brown \citep{Hoek1980} came up with a new particular shape of quadratic nonlinear criterion. It reproduced the Mohr--Coulomb type singularity for weak tensile loading, simultaneously taking into account the reduction of shear resistance for high compression. Purely empirical, the Hoek--Brown criterion was originally obtained for intact rocks by two-parameter fitting the results of triaxial tests. Validated during the following years on a wider experimental database, the criterion was used extensively in the design of underground excavations \citep{Hoek1982}, and was later extended to the 3D case in works of 
\citep{Pan1988}
. 
\\
 
In this article, we propose a way to derive a quadratic Hoek--Brown (Pan--Hudson) type yield criterion from the linear Mohr--Coulomb (Drucker--Prager) one.  The Hoek--Brown relation is seen as a consequence of the simultaneous presence of both plasticity and nonlinear hyperelasticity phenomena. 
Additionally, inspired by the foundational ideas of \citep{Dafalias1975,Vermeer1984}, we argue that the cyclic triaxial compression test plays a distinctive role in the material classification, as it reveals the potential presence of hysteretic nonlinearity. Building on the same key insight that nonlinearity should occur inside some fixed ultimate yield surface (called bounding surface in  \citep{Dafalias1975}), and not only on its boundary as in classical plasticity, we develop a perfect plasticity model that incorporates hyperelastic nonlinearity as a core feature. This simplest model has an entirely analytical numerical integration scheme, paving the way for its further refinement and extension.\\

The paper is organized as follows. In \Cref{sec:hyperelasticity coupled to plasticity}, 
we outline first the main features and limitations of a linear elasto-plastic model, and then we discuss some choices for the extension to a nonlinear elasto-plastic model. 
\Cref{sec:hyperelastic model} is dedicated to the description of a hyperbolic elasticity in combination of the Drucker--Prager yield criterion, which represents the main focus of the paper. 
In \Cref{sec:numerical}, we analyze the response of this model with some typical tests for a material point. In particular, we show the saturation of dilatancy and the accommodation phenomenon during triaxial cyclic loadings, and we also compare the proposed nonlinear model with the linear elasto-plastic one. 
An experimental comparison with a rock on a uniaxial compression test is then discussed
. In \Cref{sec:Goustan}, we present structural finite element simulations demonstrating the numerical efficiency of the model.
Finally, in \ref{sec:HB_history} and 
in \ref{sec:numerical_integration}, we present a short discussion about the history of quadratic yield criteria and we detail the numerical integration procedure of the proposed model.

\subsection{Notations}

Throughout the paper, we will use the following notation.
The symbol $\coloneqq$ denotes a definition. 
We use the usual notation of mechanics: $\strainb\in\MAT{R}^{3\times 3}_{\text{sym}}$ is the classical infinitesimal strain tensor, defined as the symmetric gradient of the displacement vector field $\VEC{u}$; $\plb\in\MAT{R}^{3\times 3}_{\text{sym}}$ is the plasticity component of the strain tensor; $\stressb \in \MAT{R}^{3\times 3}_{\text{sym}}$ is the stress tensor; $\plforb \in \MAT{R}^{3\times 3}_{\text{sym}}$ is the thermodynamical (dissipative) force associated with plasticity. \\

We adopt the usual convention in Mechanics of Continuous Media for the sign of strain and stress: the stress is positive in traction and negative in compression. The notation $\idtwo$ indicates the second-order identity tensor.
Additionally, for any symmetric fourth-order tensor $\MAT{A}$ and any second-order tensor $\VEC{\tau}$, their contraction is implicitly resulting in the second-order tensor $(\MAT{A}:\VEC{\tau})_{ij}\coloneqq A_{ijkl}\tau_{kl}$. 
The double dot is used as a symbol for the contraction of two second-order tensors: for any couple of symmetric second-order tensors $\VEC{\tau},\VEC{\eta}$, it results in the scalar expression $\VEC{\tau} : \VEC{\eta}\coloneqq  \tr (\VEC{\tau} \cdot \VEC{\eta}) = \tau_{ij}\eta_{ji}$, where $\tr$ states for the second-order tensor trace operation. 
The symbol $\MAT{I}$ is used to denote the fourth order identity tensor, and the components of the fourth order tensor resulting from a dyadic product $\VEC{\tau}\otimes\VEC{\eta}$ are expressed by $\tau_{ij}\eta_{kl}$. 
The isotropic fourth-order elasticity tensor $\MAT{E}$ is described with the help of the bulk and shear moduli, $\kappa$ and $\mu$ respectively. We also use the second-order tensor decomposition
\begin{equation}\label{eq:decomposition spherical and deviatoric}
\VEC{\tau} = \tau_m\, \idtwo + \VEC{\tau}^D, \quad \text{with} \ \tau_m = \frac{\tr\VEC{\tau}}{3},
\end{equation}
where $\tau_m$ and $\VEC{\tau}^D$ are the spherical and deviatoric parts of the tensor $\VEC{\tau}$, respectively, and we introduce the scalar norm $\left\lVert \VEC{\tau}\right\rVert \coloneqq \sqrt{\VEC{\tau} : \VEC{\tau}}$. Finally, $\MAT{K}_{\stressb}$ and $\MAT{K}_{\plforb}$ denote the elasticity domain in the stress, $\stressb$, and the thermodynamical force associated with plasticity, ${\plforb}$, space, respectively.

\section{Hyperelasticity coupled to plasticity}\label{sec:hyperelasticity coupled to plasticity}

In this Section, the main concepts of the model's formulation are summarized, the model itself being presented in \Cref{sec:hyperelastic model}.
The present objective is to establish isothermal isotropic constitutive relations for rock-like materials satisfying some basic thermodynamic principles~\footnote{More detailed descriptions of the formalism and notations can be found in classical books \citep{Simo1998, Suquet2003, Lubliner2008}.}.

\subsection{Thermodynamics of perfect plasticity}

The thermodynamical description of Continuum Mechanics relies equivalently on the Helmholtz free energy \citep{Germain1983} or alternatively on Gibbs free energy \citep{Collins1997,Houlsby2007} density function. The existence of such an energy function state may be derived from the more basic Work Principle introduced in \citep{Marigo1989} that appears naturally for any continuum mechanical system. For isothermal evolution in the presence of plasticity, this function depends on at least two state variables. We make the choice of Helmholtz description with the total infinitesimal strain $\strainb$ and the plastic strain $\plb$ as such state variables, so that $ \psi(\strainb,\plb)$ is a scalar function representing the free energy of the system. \\

As we have mentioned in the introduction, considerable experimental efforts have been focused on the identification of the elasticity domain. In the perfect plasticity hypothesis, the domain is supposed to be fixed during evolution, and its shape can consequently be considered as one of the main material properties. Following original Green-Naghdi ideas, we admit then the existence of an elasticity domain $\MAT{K}_{\stressb}$ in the stress space. \\

Thermodynamically, one usually needs to ensure that the intrinsic dissipation $\cal D$, alternatively called interior work \citep{Marigo1989}, whenever hardening is ignored in the free energy, to be positive:
\begin{equation}\label{eq:dissipation}
{\cal D} =\stressb : \dot{\strainb} - \dot\psi (\strainb,\plb)\ge 0,
\end{equation}
where the dot denotes the temporal derivative.
Considering elastic unloading first, we obtain classically the relation $\stressb  = \partial\psi/\partial\strainb$ as the consequence of a zero dissipation process, ``${\cal D(\plb=\text{const})}=0$'', reducing then the second thermodynamic principle in \cref{eq:dissipation} to the straight inequality
\begin{equation}\label{eq:dissipation_simple}
{\cal D} = -\frac{\partial \psi}{\partial \plb} :\dot{\plb} = \plforb : \dot{\plb}\ge 0.
\end{equation}

The thermodynamical force associated with plasticity (\emph{plastic force}), $\plforb = -\partial\psi/\partial\plb$, captures the dissipation of the system in parallel with the plastic strain evolution. \\

The simplest elasto-plastic models admit additive separation of elastic and plastic strains, resulting in the quadratic expression for the free energy
\begin{equation}\label{eq:scalar_energy density}
\psi(\strainb,\plb) = \frac{1}{2} (\strainb-\plb) :\MAT{E}:(\strainb-\plb).
\end{equation}
We recover thus the classical Hooke's law with some residual plasticity
\begin{equation}
\stressb = \frac{\partial \psi}{\partial\strainb}= \MAT{E}:(\strainb - \plb).
\end{equation}
Additionally, the plastic force and the stress expression coincide, since
\begin{equation}\label{eq:sigma equal X}
\plforb= -\frac{\partial\psi}{\partial\plb} =  \MAT{E}:(\strainb - \plb) =\stressb.
\end{equation}

It is then equivalent to define the elasticity domain either in the stress space or in the plastic force space. Nevertheless, for some general expressions of the free energy $\psi(\strainb,\plb)$, the stress and the plastic force are different physical quantities. Even if for experimental scientists, it is undoubtedly more convenient to analyze the elasticity domain shape through the observable stress tensor, the portrayal with the plastic force benefits from one conceptual advantage in thermodynamics, as the plastic force is, by \cref{eq:dissipation_simple}, the 
correct thermodynamical force 
to govern the intrinsic dissipation during the irreversible processes with plastic evolution. \\

Furthermore, the positiveness of the intrinsic dissipation (${\cal D}\geq 0$) can be advantageously systematically satisfied with the help of the so-called normality flow rule on the plastic strain, which is one consequence of using the Generalized Standard Materials (GSM) framework \citep{Halphen1975}\footnote{See also for instance \citep{Ortiz1999} for the benefit of the GSM framework to solve constitutive equations by the use of incremental variational formulations.}. In this context, the intrinsic dissipation relates to a combination of the geometrical properties of the elasticity domain shape.

\subsection{Beyond the linear approximation}\label{subsec:line_approx}

Most materials exhibit an initial linear elastic phase before developing a more complex nonlinearity. The origin of nonlinearity may be purely geometrical, captured through finite strain description, but for some materials, it is their specific microstructure that results in a macroscopic reversible nonlinear behavior. Processes such as void compression in sands, micro-cracking closure in concrete and rocks, and void creation or collapse in some metals, all contribute to the effective homogenized nonlinear stress-strain dependence, which may be captured through an infinitesimal strain description. We highlight a few historically influential models that were introduced across diverse application domains. The empirical three-parameter fitting stress-strain relation proposed by Ramberg and Osgood \citep{Ramberg1943} was initially introduced for more accurate modeling of alloys, but is still widely used for metals under fracture conditions. In the late seventies, the analysis of experimental data from the constrained biaxial loading of concrete, known as the Kupfer test \citep{Kupfer1973}, sparked a debate about the origin and significance of the elastic coefficients' dependence on loading strength. Addressing the same issue and following the original ideas of Ambarsumyan, Curnier  \citep{Curnier1994} introduced a tension/compression loading asymmetry in materials using an energy-based, cone-wise elastic constitutive relation, which is found to be suitable for the description of partially damaged materials \citep{Vicentini2024}. In \citep{Houlsby1985, Niemunis1998} the authors mapped multiple constitutive relations for clay into a hyperelastic formalism, 
aiming to satisfy dissipation-free evolutions
. This approach is gaining acceptance in the soil community with a progressive replacement of the incremental hypoelasticity. Although the motivation and inherent mechanical explanations for introducing nonlinear elastic stress-strain relations have varied along time, 
ranging from experimental observations to micromechanical insights \citep{Hicher1996, Maalej2007}
, it has been proven to be a highly relevant tool for material modeling, especially in the case of rock and soil. In this Section, we will show how nonlinear hyperelasticity fits into the GSM formalism, resulting to a nonlinear relationship between $\stressb$  and $\plforb$. \\

The nonlinear hyperelasticity is introduced in a phenomenological way, by defining a Helmholtz free energy $\psi(\strainb,\plb)$ that generates the in-equivalence $\stressb\neq\plforb$.  
This approach deviates from the more conventional method where nonlinearity is introduced solely through the elastic strain \citep{Houlsby1985,Borja_1997,Bacquaert2024}.
\\

We consider here one possible extension, where the isotropic elastic moduli $\kappa(\strainb)$ and $\mu(\strainb)$ are supposed to be some functions of the total strain $\strainb$, through
\begin{align}\label{eq:scalar_energy density non lin}
\begin{split}
\psi(\strainb,\plb) \coloneqq&\ \frac{1}{2} (\strainb-\plb):\MAT{E}(\strainb):(\strainb-\plb) \\ 
=&\ \frac{\kappa(\strainb)}{2}[\tr(\strainb-\plb)]^2 + \mu(\strainb)\, (\strainb^D-\plb^D):(\strainb^D-\plb^D).
\end{split}
\end{align}

In this case, the stress-strain relationship becomes
\begin{align}\label{eq:scalar_nl_stress1}
\begin{split}
\stressb &= \frac{\partial\psi (\strainb,\plb)}{\partial \strainb} \\
&= \MAT{E}(\strainb):(\strainb - \plb) + \frac{1}{2} \frac{\partial\kappa(\strainb)}{\partial \strainb}[\tr(\strainb-\plb)]^2 + \frac{\partial\mu(\strainb)}{\partial \strainb}(\strainb^D-\plb^D) : (\strainb^D-\plb^D),
\end{split}
\end{align}
which has a clear nonlinear contribution\footnote{Note that the partial derivatives ${\partial\kappa(\strainb)}/{\partial \strainb}$ and ${\partial\mu(\strainb)}/{\partial \strainb}$ in \cref{eq:scalar_nl_stress1} are symmetric second-order tensors.}, while the plastic force preserves the standard expression
\begin{equation}\label{eq:Xp non linear}
\plforb =-\frac{\partial\psi (\strainb,\plb)}{\partial \plb}= \MAT{E}(\strainb):(\strainb - \plb) = \kappa(\strainb)\tr(\strainb-\plb)\idtwo + 2\mu(\strainb)\, (\strainb^D-\plb^D).
\end{equation}
Accordingly, the stress is no longer equal to the plastic force. \\ 

Combining \cref{eq:scalar_nl_stress1} and \cref{eq:Xp non linear}, the plastic strain can be eliminated in the stress-strain relationship to obtain
\begin{equation}\label{eq:scalar_nl_stress2}
 \stressb = \plforb + \frac{1}{2\kappa(\strainb)^2}\frac{\partial\kappa(\strainb)}{\partial \strainb}\plform^2 + \frac{1}{4\mu(\strainb)^2}\frac{\partial\mu(\strainb)}{\partial \strainb} \plforb^D :\plforb^D.
\end{equation}

This equation represents the nonlinear \emph{a priori} strain-dependent mapping that transforms any domain defined in the plastic force space into the stress space. Therefore, in the present proposed model, where the elasticity domain is fixed in the plastic force space, the presence of nonlinear hyperelasticity introduces not only a nonlinear stress-strain relationship, but also affects the elasticity domain shape in the stress space, making it possibly strain-dependent. The nonlinear hyperelasticity is seen as a natural enrichment for the whole set of loading, including irreversible responses, and not only for reversible ones. \\

\section{Hyperelastic enrichment of perfect plasticity}\label{sec:hyperelastic model}

In this Section, we show that in the presence of nonlinear hyperelasticity, a quadratic yield criterion of Hoek--Brown type (see \ref{sec:HB_history}) can be constructed from a linear elasto-plastic model with a Drucker--Prager yield criterion.

\subsection{Mapping thermodynamics to observables}\label{subsec:energy density}

In the full generality, identifying the isotropic elastic moduli in \cref{eq:scalar_energy density non lin}, experimentally, seems to be a very difficult challenge. In the case of perfect plasticity, there exists however a particular class of hyperelasticity that preserves the non-evolving nature of the yield criterion in the stress space, which can be interpreted either as an initial yield criterion or as a residual one. We show hereafter that this condition can be reached by setting both isotropic elastic moduli as a hyperbolic function of the strain trace, see \cref{eq3:compressibility and shear moduli}.
\\

For any scalar function $f(\strainb)\in\mathbb{R}$ depending on the strain only through the two first invariants $\tr\strainb$ and $\| \strainb^D \|$, the gradient simplifies to 
\begin{equation}\label{eq:scalar_chain_deriv}
\frac{\partial f(\tr\strainb, \| \strainb^D \|)}{\partial \strainb}=\frac{\partial f}{\partial \tr\strainb}\frac{\partial \tr\strainb}{\partial \strainb} +  \frac{\partial f}{\partial \| \strainb^D \|}\frac{\partial \| \strainb^D \|}{\partial \strainb} = \frac{\partial f}{\partial \tr\strainb}\idtwo +  \frac{\partial f}{\partial \| \strainb^D \|}\frac{\strainb^D}{\| \strainb^D \|}.
\end{equation}

While this expression would be derivable at $\strainb^D=\VEC{0}$ for a $\| \strainb^D \|^2$ type relation, the only expression of the function $f(\strainb)$ achieving a strain-independent gradient is deviator-independent and linear in the trace: $f(\strainb) = C_1 + C_2\tr\strainb$, with $C_1, C_2$ being two arbitrary scalar constants. If we look back into \cref{eq:scalar_nl_stress2}, where two similar second-order tensors appear as $\displaystyle\frac{1}{\kappa(\strainb)^2}\frac{\partial\kappa(\strainb)}{\partial \strainb} = -\frac{\partial\kappa^{-1}(\strainb)}{\partial \strainb}$ and $\displaystyle\frac{1}{\mu(\strain)^2}\displaystyle\frac{\partial\mu(\strainb)}{\partial \strainb} = -\frac{\partial\mu^{-1}(\strainb)}{\partial \strainb}$, the same strain-independent condition on their gradients implies a hyperbolic dependence of the strain trace for the two isotropic elastic moduli introduced in the free energy \cref{eq:scalar_energy density non lin}:
\begin{equation}\label{eq3:compressibility and shear moduli}
\kappa(\strainb) = \frac{\kappa_i}{1+2\kappa_i \beta_m\tr\strainb},\quad 
\mu(\strainb) = \frac{\mu_i}{1+4\mu_i\beta^D\tr\strainb},
\end{equation}
where $\kappa_i$ and $\mu_i$ are the initial compressibility and shear moduli for $\tr\strainb=0$, $\beta_m \geq 0$ and $\beta^D \geq 0$ are the nonlinear parameters (both homogeneous to compliance). For simplicity, however, only the case $\beta_m > 0$ will be discussed in the following, whereas linear shear elasticity will be conserved with $\beta^D=0$. 

\begin{remark}
Another compelling case would arise by considering $\beta^D> 0$. This would result in a hyperelastic deviatoric-volumetric coupling as it has been observed and modeled in the literature \citep{Houlsby1985, Borja_1997, Houlsby_2019}. This represents a promising direction for further investigation, as outlined in the perspectives section.
\end{remark}

Following \eqref{eq:scalar_nl_stress1}, \eqref{eq:Xp non linear} and \eqref{eq3:compressibility and shear moduli}, we obtain
\begin{equation}\label{eq:stress expression new}
\stressb = \sigma_m\idtwo + \stressb^D,\quad\text{with}\quad 
\begin{cases}
\begin{aligned}
\stress_m 
&=\cfrac{\kappa_i}{1+2\kappa_i\beta_m\tr\strainb} \tr(\strainb -\plb) \\
&-\cfrac{\kappa_i^2\beta_m}{(1+2\kappa_i\beta_m\tr\strainb )^2} [\tr(\strainb -\plb)]^2 \\
\end{aligned}\vspace{4mm} \\
\stressb^D 
= 2\mu_i(\strainb^D-\plb^D),
\end{cases}
\end{equation}
and
\begin{equation}\label{eq:X expression new}
\plforb = \plform\idtwo + \plforb^D,\quad\text{with}\quad 
\begin{cases}
\plform 
=\cfrac{\kappa_i}{1+2\kappa_i \beta_m \tr\strainb}\tr(\strainb -\plb)  \vspace{4mm} \\
\plforb^D 
= 2\mu_i (\strainb^D-\plb^D).
\end{cases}
\end{equation}

As a result, $\stressb$ and $\plforb$ are connected through the strain-independent relationship
\begin{equation}\label{eq:relation between stress and X new}
\stressb = \plforb - \beta_m\plform^2\idtwo.
\end{equation}

One can observe that the hyperbolic elasticity \cref{eq3:compressibility and shear moduli} generates a quadratic relationship between the spherical parts of the stress $\stressb$ and the plastic force $\plforb$. \\

Once again, as previously pointed out in the previous \Cref{sec:hyperelasticity coupled to plasticity}, the nonlinear relation \cref{eq:relation between stress and X new} can be viewed as a transformation rule from the thermodynamically defined elasticity domain (in the space of the plastic force) to the experimentally observable elasticity domain (in the space of the stress). Specifically, any shape in the plastic force $\plforb$ (the thermodynamic one) has a one-to-one map to its counterpart in the observable stress space $\stressb$ (the experimental one).  While this opens up a broad area for investigation, our focus in this paper remains primarily on the simplest example of transforming the Drucker--Prager yield criterion, as detailed in the following.

\subsection{The yield criterion transformation}\label{sec:domain transformation}

We consider the transformation of the Drucker--Prager yield criterion by the hyperbolic elasticity. In the space of the plastic force, the criterion reads
\begin{equation}\label{eq:Drucker-Prager criterion X}
f_{\plforb}(\plforb) \coloneqq \frac{1}{\sqrt{6}}\left\lVert\plforb^D\right\rVert + a \plform - b,
\end{equation}
where $a$ and $b$ stand as friction and cohesion parameters, respectively.
The corresponding elasticity domain $\MAT{K}_{\plforb} \coloneqq \{ \plforb^*\in\MAT{R}^{3\times 3}_{\rm sym} \, |\, f_{\plforb}(\plforb^*) \leq 0 \}$ is a convex cone with linear boundary and a singular point at
$(\plform,\plforb^D)=(b/a, \VEC{0})$. 
Moreover, following the GSM framework, the normality flow rule is adopted. In the points in which the boundary is smooth, \emph{i.e.}, where the function $f_{\plforb}$ is differentiable, it leads to 
\begin{equation}\label{eq:flow rule X}
\dot{\plb} = \dot{\lambda} \left(\frac{1}{\sqrt{6}} \frac{\plforb^D}{\lVert\plforb^D\rVert} + \frac{a}{3}\, \idtwo \right),
\end{equation}
where the plastic multiplier $\dot{\lambda}$ verifies the consistency conditions
\begin{equation}
     \dot{\lambda} \geq 0,\quad f_{\plforb}(\plforb) \leq 0, \quad \dot{\lambda}f_{\plforb}(\plforb)=0.
\end{equation}

\begin{figure}[ht]
\centering
\includegraphics{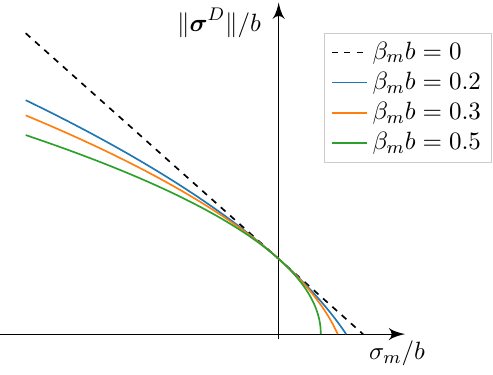}
\caption{Transformation of the yield criterion for different values of the nonlinear parameter $\beta_m$. The black dashed line represents the case of the linear Drucker--Prager yield criterion for $\beta_m=0$. 
}
\label{fig:Domain transformation tensor}
\end{figure}

By combining \cref{eq:Drucker-Prager criterion X} with \cref{eq:relation between stress and X new}, one can express the explicit expression of the yield criterion in the stress space as
\begin{equation}\label{eq:criterion in stress space}
f_{\stressb}(\stressb) = \frac{\beta_m}{6}\left\lVert\stressb^D\right\rVert^2 + \frac{a-2\beta_mb}{\sqrt{6}} \left\lVert\stressb^D\right\rVert + a^2\stress_m -  b\left(a-\beta_m b\right).
\end{equation}
Therefore, starting from the linear yield criterion \cref{eq:Drucker-Prager criterion X} for the plastic force $\plforb$, a quadratic yield criterion for the stress $\stressb$ is obtained. The corresponding elasticity domain $\MAT{K}_{\stressb} \coloneqq \{ \stressb^*\in\MAT{R}^{3\times 3}_{\rm sym} \, |\, f_{\stressb}(\stressb^*) \leq 0 \}$ is a convex cone with parabolic boundary and a singular point at $(\stress_m,\stressb^D)=(b(a-\beta_m b)/a^2, \VEC{0})$. \Cref{fig:Domain transformation tensor} shows the transformation of the yield criterion for different values of the nonlinear parameter $\beta_m$. In this Figure, $a=1$. 


Furthermore, if the conditions $a-2\beta_mb\geq 0$ and $b\geq 0$ hold, \cref{eq:criterion in stress space} shows that the domain in the stress space is convex and contains the origin. As well, the second invariant of the stress can be written as
\begin{equation}
\frac{1}{\sqrt{6}}\left\lVert\stressb^D\right\rVert = \frac{-(a-2\beta_mb) + a\sqrt{1-4\stress_m\beta_m}}{2\beta_m}.
\end{equation}
The Drucker--Prager expression is naturally recovered for vanishing nonlinear hyperelasticity, \emph{i.e.}, with $\beta_m=0$,
\begin{equation} \label{eq:Drucker-Prager criterion sigma}
    \frac{1}{\sqrt{6}}\left\lVert\stressb^D\right\rVert = b-a\stress_m.
\end{equation}

To summarize this Section, we have shown a possible transformation of the linear Drucker--Prager cone to a quadratic Hoek--Brown one, generated by the hyperbolic elasticity. The \Cref{tab:summary_equations} gathers all the constitutive equations of the proposed nonlinear elasto-plastic model.

\begin{table}[ht]\footnotesize
\begin{center}
\begin{tabular}{l}
\hline
\textbf{The state variables:}\\
\hspace{80pt}the total infinitesimal strain $\strainb$ \\
\hspace{80pt}the plastic strain $\plb$ \\[3ex]
\textbf{The free energy:}\\
\hspace{80pt} $\psi(\strainb,\plb) := \cfrac{1}{2}\cfrac{\kappa_i}{1+2\kappa_i \beta_m \tr\strainb}[\tr(\strainb -\plb)]^2+\mu_i(\strainb^D-\plb^D):(\strainb^D-\plb^D)$\\[3ex]
\textbf{The stress--strain relationship:}\\
\hspace{80pt} $\stressb = \cfrac{\kappa_i}{1+2\kappa_i\beta_m\tr\strainb} \tr(\strainb -\plb)\left(1-\cfrac{\kappa_i\beta_m}{1+2\kappa_i\beta_m\tr\strainb }\tr(\strainb -\plb)\right)\idtwo+2\mu_i(\strainb^D-\plb^D)$\\[3ex]
\textbf{The plasticity thermodynamical force:}\\
\hspace{80pt} $\plforb = \cfrac{\kappa_i}{1+2\kappa_i\beta_m\tr\strainb} \tr(\strainb -\plb)\idtwo+2\mu_i(\strainb^D-\plb^D)$\\[3ex]
\textbf{The plasticity evolution law:}\\
\hspace{80pt}the yield criterion in the plastic force space: $f_{\plforb}(\plforb) \coloneqq \cfrac{1}{\sqrt{6}}\left\lVert\plforb^D\right\rVert + a \plform - b$ \\
\hspace{80pt}the flow rule: $\dot{\plb} = \dot{\lambda}\cfrac{df_{\plforb}}{d\plforb}(\plforb),\quad \dot{\lambda} \geq 0,\quad f_{\plforb}(\plforb) \leq 0, \quad \dot{\lambda}f_{\plforb}(\plforb)=0$\\[3ex]
\textbf{The yield criterion in the stress space:}\\
\hspace{80pt} $f_{\stressb}(\stressb) = \cfrac{\beta_m}{6}\left\lVert\stressb^D\right\rVert^2 + \cfrac{a-2\beta_mb}{\sqrt{6}} \left\lVert\stressb^D\right\rVert + a^2\stress_m -  b\left(a-\beta_m b\right)$
\\ [1ex] 
 \hline
\end{tabular}
\end{center}
\caption{Summary of the constitutive equations of the nonlinear elasto-plastic model.}
\label{tab:summary_equations}
\end{table}

\begin{remark}
It can be easily shown that the free energy $\psi(\strainb,\plb)$ in \Cref{tab:summary_equations} is a strictly convex function of $\strainb$ and $\plb$ separately, and that the determinant of its Hessian matrix is zero.
\end{remark} 

\section{Responses on typical tests}\label{sec:numerical}

In this Section, we present a panel of examples of evolution using the proposed nonlinear elasto-plastic model in some typical test cases. In particular, the influence of the nonlinear elastic parameter $\beta_m$ is exhibited, and a comparison with the linear elasto-plastic model is provided. Loading conditions focus on hydrostatic and triaxial (single compression and cyclic) tests. The results have been obtained by solving the constitutive equations of the model with the open source code generation tool MFront \citep{mfront}. 
The numerical integration procedure is summarized in \ref{sec:numerical_integration}
.\\

For all the following tests, both initial bulk and shear moduli are set as $\kappa_i = 5E/6$ and $\mu_i = 5E/13$, where the Young modulus is $E=\SI{100}{\mega\pascal}$, and the ratio $\kappa_i/\mu_i$ is equivalent to a Poisson's ratio of $0.3$. The friction and cohesion parameters are $a=1/9$ and $b=E/3000$.  The nonlinear hyperelastic parameter is  $\beta_m=120/E$.

\subsection{Hydrostatic tests}\label{sec:hydro_test}

In the hydrostatic test, the stress remains on the hydrostatic axis throughout the loading.  Only the spherical part of the stress evolves, \emph{i.e.}, $\stressb = \bar{\stress} \idtwo$, with $\bar{\stress} < 0$ for compression, and $\bar{\stress} > 0$ for traction. \\

By using \cref{eq:stress expression new} in the initial elastic regime ($\plb = \VEC{0}$), we have immediately $\strainb^D = \VEC{0}$ and the simple hyperbolic relation
\begin{equation}\label{eq:evolution stress hydrostatic}
\bar{\stress} = \frac{1}{4 \beta_m} \left(1 - \frac{1}{(1+2\kappa_i\beta_m \tr\strainb)^2} \right),
\end{equation}
or equivalently
\begin{equation}\label{eq:evolution strain hydrostatic}
\tr\strainb = \frac{1}{2\kappa_i\beta_m} \left( \frac{1}{\sqrt{1 - 4\beta_m \bar{\stress}}} - 1 \right).
\end{equation}

\begin{figure}[ht]
\begin{subfigure}{0.49\textwidth}
\centering
\resizebox{!}{0.7\textwidth}{
\includegraphics{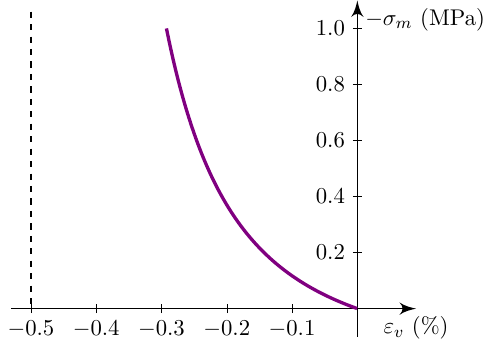}
}
\caption{Compression}
\label{fig:hydrostatic triaxial test compression}
\end{subfigure}
\hfill
\begin{subfigure}{0.49\textwidth}   
\centering
\resizebox{!}{0.7\textwidth}{%
\includegraphics{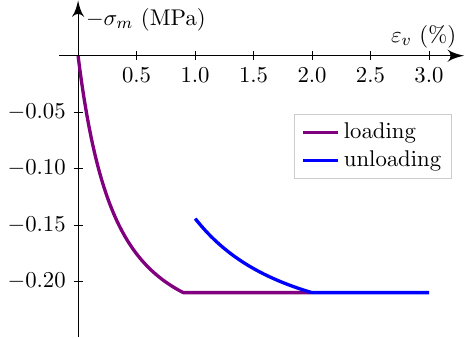}
}
\caption{Traction}
\label{fig:hydrostatic triaxial test traction}
\end{subfigure}
\caption{Graphs of hydrostatic triaxial tests. While during hydrostatic compression, loading and unloading follow the same curve, in a traction test, this is not the case due to plastification. This is highlighted with colors, with {\it violet} representing the first loading and {\it blue} representing the unloading and second loading.
}
\label{fig:hydrostatic triaxial test}
\end{figure}

According to the yield criterion \cref{eq:criterion in stress space}, this relation still holds for any value of hydrostatic compression and for the first part of hydrostatic traction. In a traction test indeed, initially the behavior is elastic until the spherical part of the stress reaches the maximum $b(a- \beta_m b)/a^2$. Then, plasticity starts. One can see the corresponding graphs of the evolution of $\bar\stress$ as a function of $\tr\strainb$ in \Cref{fig:hydrostatic triaxial test}. One can observe, in agreement with \cref{eq:evolution stress hydrostatic} for $\bar\stress \rightarrow -\infty$, that a hydrostatic compression limit $\strain_0$ is predicted by
\begin{equation}
\strain_0 = -\frac{1}{2\kappa_i\beta_m}.
\end{equation}

\subsection{Triaxial compression test with a confining pressure}\label{sec:triaxial compression}

A triaxial compression test with a confining pressure is divided into two stages: first, a hydrostatic compression to reach a confining pressure $p_0$, then an axial compression while maintaining the lateral pressure at $p_0$. \\

As it has been shown for the hydrostatic tests, at the end of the first stage, $\stressb = -p_0\idtwo$ and $\strainb = (\tr\strainb/3)\idtwo$, where the explicit expression of the volumetric strain $\tr\strainb$ has been obtained in \cref{eq:evolution strain hydrostatic}. During the second stage, the axial strain is increased (in compression), stress and strain tensors can thus be written, by symmetry, in the basis $(\VEC{e}_x,\VEC{e}_y,\VEC{e}_z)$ where $z$ denotes the axial direction and $x$ and $y$ the lateral ones, as
\begin{equation*}
    \stressb = \begin{pmatrix}
    -p_0 & 0 & 0 \\ 
    0 & -p_0 & 0 \\
    0 & 0 & \stress_z
    \end{pmatrix},\quad
    \strainb = \begin{pmatrix}
    \strain_x & 0 & 0 \\
    0 & \strain_x & 0 \\
    0 & 0 & \strain_z
    \end{pmatrix}.
\end{equation*}
The first two stress invariants are thus
\begin{equation}\label{eq:sigma_m+sigma_D}
    \stress_m = \displaystyle\frac{1}{3} (\stress_z - 2 p_0),\quad
    \lVert\stressb^D\rVert = \sqrt{\frac{2}{3}}\, \lvert q \rvert,
\end{equation}
where we note $q \coloneqq p_0 + \stress_z$ the equivalent shear stress, as usual for Soil Mechanics.
\\

\begin{figure}[ht]
    \begin{subfigure}{0.49\textwidth}
        \centering
        \resizebox{!}{0.7\textwidth}{%
            \includegraphics{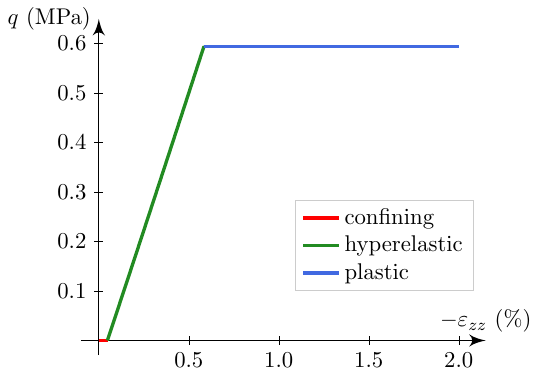}
	}
	\caption{Equivalent shear stress $q = \stress_x - \stress_z$ vs. axial strain $\strain_z$.}
	\label{fig:uniax - q vs eps_z}
    \end{subfigure}
    \hfill
    \begin{subfigure}{0.49\textwidth}   
        \centering
        \resizebox{!}{0.7\textwidth}{%
            \includegraphics{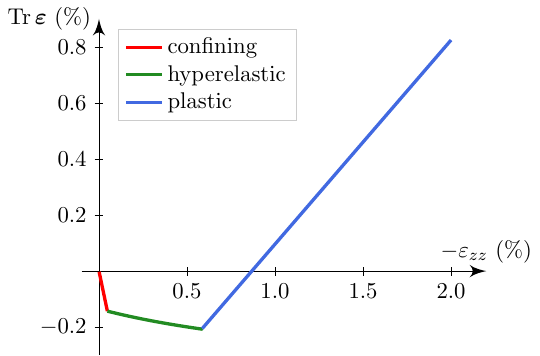}
	}
	\caption{Volumetric strain $\tr\strainb$ vs. axial strain $\strain_z$.}
	\label{fig:uniax - dil}
    \end{subfigure}
    \medskip
    \\
    \begin{subfigure}{0.49\textwidth}
        \centering
        \resizebox{!}{0.7\textwidth}{%
            \includegraphics{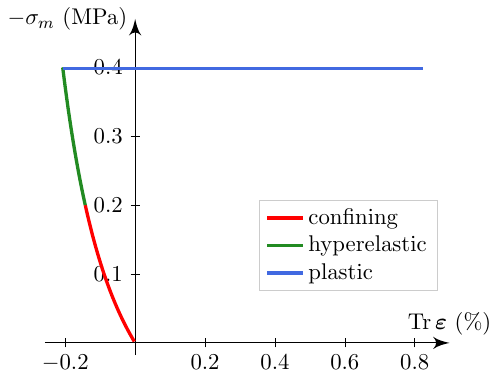}
	}
	\caption{Average stress $\stress_m$ vs. volumetric strain $\tr\strainb$.}
	\label{fig:uniax - sig_m vs tr eps}
    \end{subfigure}
    \hfill
    \begin{subfigure}{0.49\textwidth}   
        \centering
        \resizebox{!}{0.7\textwidth}{%
            \includegraphics{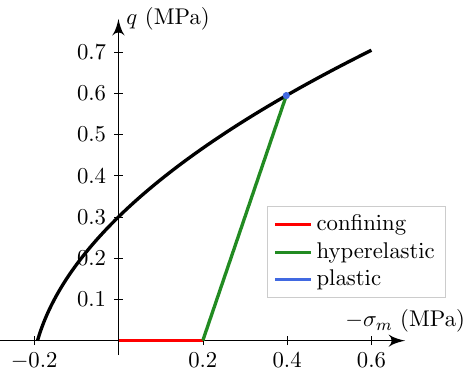}
	}
	\caption{Evolution in the stress domain.}
	\label{fig:uniax - domain}
    \end{subfigure}
    \caption{Graphs of the triaxial compression test. The different regimes of the response are shown with different colors: confining first stage (\emph{red}), elastic evolution in the second stage (\emph{green}), and plastic evolution (\emph{blue}).}
    \label{fig:triaxial compression test}
\end{figure}

\Cref{fig:triaxial compression test} shows the evolution curves for the triaxial compression test with a confining pressure of $p_0=\SI{0.2}{\mega\pascal}$. \Cref{fig:uniax - q vs eps_z,fig:uniax - dil} show the equivalent shear stress $q$ and the volumetric strain $\tr \VEC{\strain}$ as functions of the axial strain $\strain_z$, respectively, \Cref{fig:uniax - sig_m vs tr eps} shows the spherical stress $\stress_m$ as a function of the volumetric strain $\tr \VEC{\strain}$, and finally \Cref{fig:uniax - domain} shows the evolution of the stress in the elasticity domain $\MAT{K}_{\stressb}$. The different phases of the evolution are highlighted with different colors in \Cref{fig:triaxial compression test}: \emph{red} for the confining first stage, \emph{green} for the elastic evolution in the second stage, and \emph{blue} during the plastic evolution. The nonlinearity is visible during the elastic evolution (in \emph{green}) in \Cref{fig:uniax - dil,fig:uniax - sig_m vs tr eps}. During the plastic evolution, the stress lies on the boundary of the elasticity domain $\MAT{K}_{\stressb}$ and the plastic strain evolves following the normality flow rule \cref{eq:flow rule X}. The stress remains constant since the lateral pressure is maintained equal to $-p_0$ along the $x$-axis and the $y$-axis, and there is no hardening (see Figure \ref{fig:uniax - domain}).

\subsection{Cyclic triaxial test}\label{sec:cyclic}

A cyclic triaxial test is now considered. As in the \Cref{sec:triaxial compression}, a hydrostatic compression is first prescribed, until a confining pressure of $p_0=\SI{0.1}{\mega\pascal}$, before the axial strain $\strain_z$ increases and decreases (in compression) cyclically, maintaining the lateral pressure constant. \\

\begin{figure}[ht]
\begin{subfigure}{0.49\textwidth}
\centering
\resizebox{!}{0.7\textwidth}{%
\includegraphics{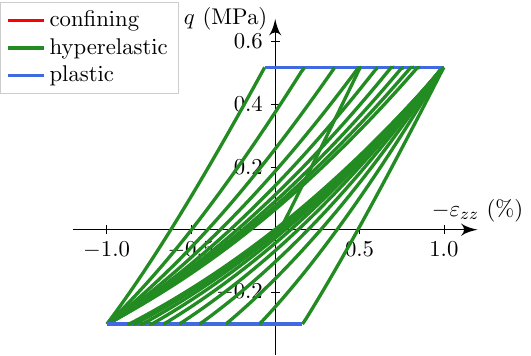}
}
\caption{Equivalent shear stress $q = \stress_x - \stress_z$ vs. axial strain $\strain_z$.}
\label{fig:cyc - q vs eps_z}
\end{subfigure}
\hfill
\begin{subfigure}{0.49\textwidth}   
\centering
\resizebox{!}{0.7\textwidth}{%
\includegraphics{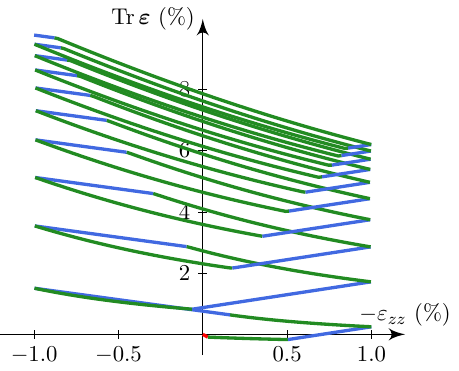}
}
\caption{Volumetric strain $\tr\strainb$ vs. axial strain $\strain_z$.}
\label{fig:cyc - dil}
\end{subfigure}
\medskip
\\
\begin{subfigure}{0.49\textwidth}
        \centering
        \resizebox{!}{0.7\textwidth}{%
            \includegraphics{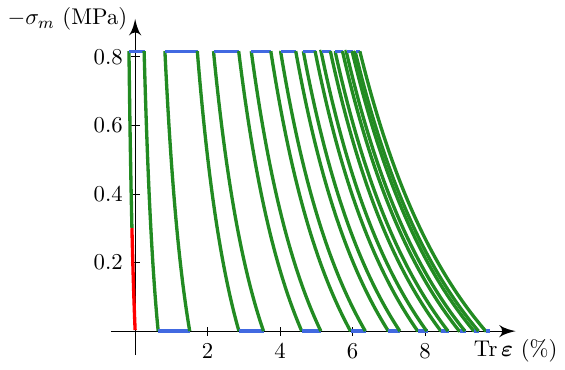}
	}
	\caption{Average stress $\stress_m$ vs. volumetric strain $\tr\strainb$.}
	\label{fig:cyc - sig_m vs tr eps}
    \end{subfigure}
    \hfill
    \begin{subfigure}{0.49\textwidth}   
        \centering
        \resizebox{!}{0.7\textwidth}{%
            \includegraphics{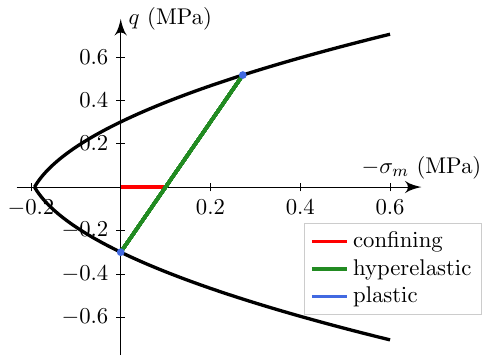}
	}
	\caption{Evolution in the stress domain.}
	\label{fig:cyc - domain}
    \end{subfigure}
    \caption{Graphs of the cyclic test.  The different regimes of the response are shown with different colors: confining first stage (\emph{red}), elastic evolution in the second stage (\emph{green}), and plastic evolution (\emph{blue}).}
    \label{fig:cyclic test}
\end{figure}

Results are displayed in \Cref{fig:cyclic test}. One can observe that the main influence of nonlinear hyperelasticity, besides the nonlinear evolution of $\tr\strainb$, $\stress_m$, and $q$, is the progressive accommodation of their values. This is particularly relevant for the saturation of the dilatancy as shown in \Cref{fig:cyc - dil}. Furthermore, \Cref{fig:cyc - domain} shows the piecewise linear evolution in the stress space. The intersections of the parabolic domain represent the stress points during the plastic phases. The dilatancy saturation, in \Cref{fig:cyc - dil} results from the progressive decrease of the isotropic bulk modulus $\kappa(\strainb)$ defined by \cref{eq3:compressibility and shear moduli}, because of positive volumetric plastic strain increment, until the response approaches a purely elastic evolution when the strain amplitude is no more enough to reach plastification. This observation is a very distinguished feature of the nonlinear elastic description, with a hyperbolic elasticity, which we have introduced in the proposed model.

\subsection{Comparison with the linear elasto-plastic model}

As it was mentioned in the \Cref{sec:cyclic}, some non-trivial results for the cyclic loading test with the current nonlinear elasto-plastic model have been shown, 
such as 
the progressive accommodation of the volumetric strain and the saturation of dilatancy. Hereunder, a comparison of the responses, with the linear elasto-plastic one, for the tests of \Cref{sec:triaxial compression} and \Cref{sec:cyclic} is briefly considered. \\

\begin{figure}[ht]
\begin{subfigure}{0.49\textwidth}
\centering
\resizebox{!}{0.7\textwidth}{%
\includegraphics{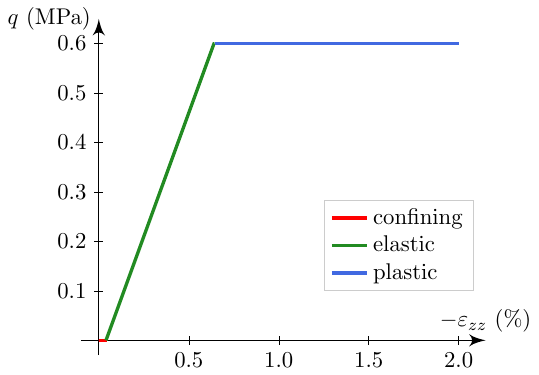}
}
\caption{Equivalent shear stress $q = \stress_x - \stress_z$ vs. axial strain $\strain_z$.}
\label{fig:comp_uniax - q vs eps_z}
\end{subfigure}
\hfill
\begin{subfigure}{0.49\textwidth}   
\centering
\resizebox{!}{0.7\textwidth}{%
\includegraphics{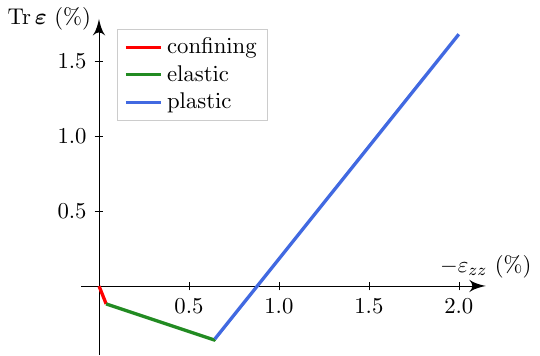}
}
\caption{Volumetric strain $\tr\strainb$ vs. axial strain $\strain_z$.}
\label{fig:comp_uniax - dil}
\end{subfigure}
\medskip
\\
\begin{subfigure}{0.49\textwidth}
\centering
\resizebox{!}{0.7\textwidth}{%
\includegraphics{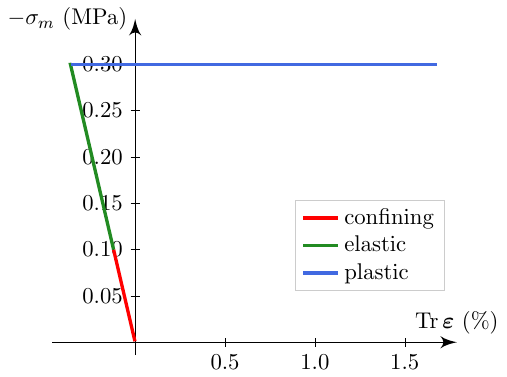}
}
\caption{Average stress $\stress_m$ vs. volumetric strain $\tr\strainb$.}
\label{fig:comp_uniax - sig_m vs tr eps}
\end{subfigure}
\hfill
\begin{subfigure}{0.49\textwidth}   
\centering
\resizebox{!}{0.7\textwidth}{%
\includegraphics{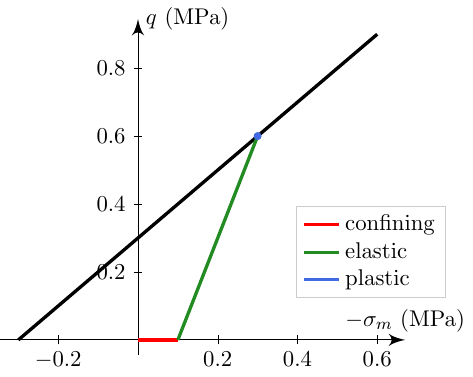}
}
\caption{Evolution in the stress domain.}
\label{fig:comp_uniax - domain}
\end{subfigure}
\caption{Response of the linear elasto-plastic model for a triaxial compression test. The different regimes of the response are shown with different colors: confining first stage (\emph{red}), elastic evolution in the second stage (\emph{green}), and plastic evolution (\emph{blue}).}
\label{fig:comp triaxial compression test}
\end{figure}

\begin{figure}[ht]
\begin{subfigure}{0.49\textwidth}
\centering
\resizebox{!}{0.7\textwidth}{%
\includegraphics{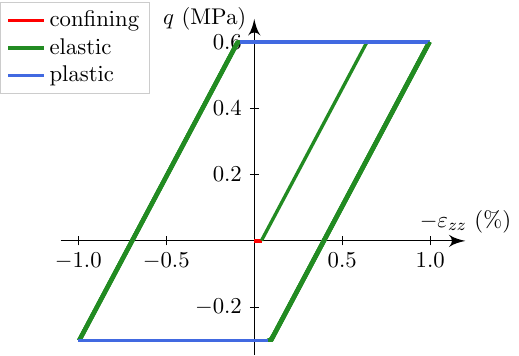}
}
\caption{Equivalent shear stress $q = \stress_x - \stress_z$ vs. axial strain $\strain_z$.}
\label{fig:comp_cyc - q vs eps_z}
\end{subfigure}
\hfill
\begin{subfigure}{0.49\textwidth}   
\centering
\resizebox{!}{0.7\textwidth}{%
\includegraphics{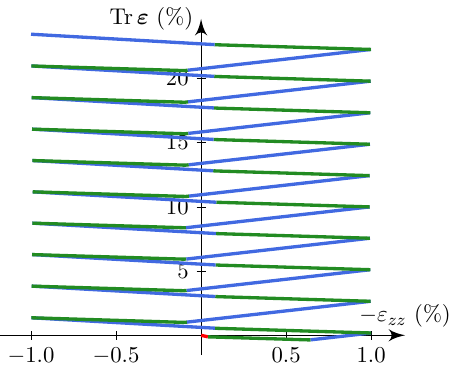}
}
\caption{Volumetric strain $\tr\strainb$ vs. axial strain $\strain_z$.}
\label{fig:comp_cyc - dil}
\end{subfigure}
\medskip
\\
\begin{subfigure}{0.49\textwidth}
\centering
\resizebox{!}{0.7\textwidth}{%
\includegraphics{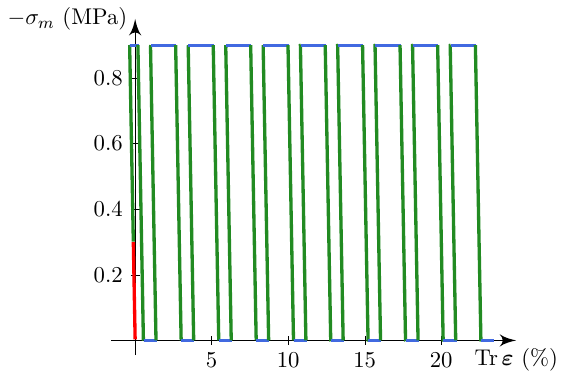}
}
\caption{Average stress $\stress_m$ vs. volumetric strain $\tr\strainb$.}
\label{fig:comp_cyc - sig_m vs tr eps}
\end{subfigure}
\hfill
\begin{subfigure}{0.49\textwidth}   
\centering
\resizebox{!}{0.7\textwidth}{%
\includegraphics{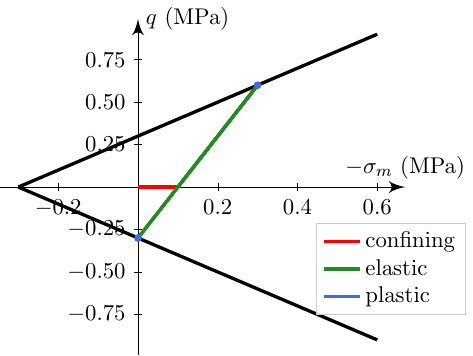}
}
\caption{Evolution in the stress domain.}
\label{fig:comp_cyc - domain}
\end{subfigure}
\caption{Response of the linear elasto-plastic model for a cyclic test. The different regimes of the response are shown with different colors: confining first stage (\emph{red}), elastic evolution in the second stage (\emph{green}), and plastic evolution (\emph{blue}).}
\label{fig:comp cyclic test}
\end{figure}

\Cref{fig:comp triaxial compression test,fig:comp cyclic test} display the results of monotonic and cyclic triaxial tests, respectively. For both loadings, the evolution of the stress in the elasticity domain lies on a straight line (\Cref{fig:comp_uniax - domain,fig:comp_cyc - domain}). From \Cref{fig:comp_uniax - dil,fig:comp_uniax - sig_m vs tr eps}, it is evident that the linear elasto-plastic model provides a linear evolution of the volumetric strain with the axial strain. Furthermore, the slope of dilatancy evolution is larger than with the nonlinear model, as observed in \Cref{fig:uniax - dil,fig:comp_uniax - dil}. Finally, \Cref{fig:comp_cyc - dil,fig:comp_cyc - sig_m vs tr eps} show that the accommodation of the volumetric strain cannot be achieved with the linear elasto-plastic model.

\clearpage

\subsection{Comparison with an experimental uniaxial compression test}

\begin{figure}[ht]
    \begin{subfigure}{0.49\textwidth}
        \centering
        \resizebox{!}{0.7\textwidth}{%
            \includegraphics{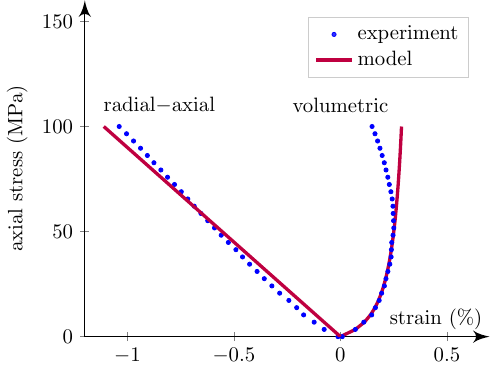}
	}
	\caption{Calibration of the elastic parameters on the initial evolution of the deviatoric and volumetric strains.}
	\label{fig:comparison_experimental_calibration}
    \end{subfigure}
    \hfill
    \begin{subfigure}{0.49\textwidth}   
        \centering
        \resizebox{!}{0.7\textwidth}{%
            \includegraphics{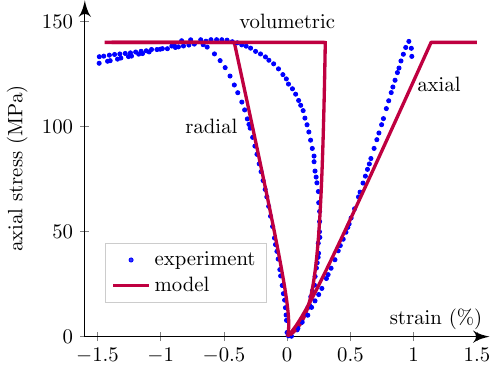}
	}
	\caption{Comparison of the proposed model on the full response of the rock material.}
	\label{fig:comparison_experimental_total}
    \end{subfigure}
    \caption{Comparison between the proposed nonlinear elasto-plastic model and experimental data from a uniaxial compression test on alkali basalt rock \citep{Heap2009}. Solid lines represent the model predictions, while dots correspond to experimental measurements. For clarity, both stress and strain are plotted as positive in compression to align with the experimental convention. The calibrated model parameters are: $\kappa_i=\SI{1.2}{\giga\pascal}, \beta_m=130/\kappa_i, \mu_i=\SI{4.5}{\giga\pascal}, a=2.8$ and $b=0$.}
    \label{fig:comparison_experimental}
\end{figure}

In this section, we present 
a short comparison between the proposed nonlinear elasto-plastic model and experimental data from a uniaxial compression test on alkali basalt rock \citep{Heap2009} (which is a triaxial compression test without confining pressure). To this end, we first calibrate the model’s elastic parameters. This process is illustrated in \Cref{fig:comparison_experimental_calibration} by fitting the initial evolution of the axial stress with the deviatoric strain, and with the volumetric strain, whose relationships involve $\mu_i$ and $(\kappa_i,\beta_m)$ respectively. For the Drucker–Prager yield criterion, we simplify the formulation by assuming $b=0$ (no cohesion). The parameter $a$ is then estimated by aligning the yield stress with the experimentally observed failure stress of approximately \SI{140}{\mega\pascal} \citep{Heap2009}. The resulting parameter values are summarized in the legend of \Cref{fig:comparison_experimental}. \\

In \Cref{fig:comparison_experimental_total}, we compare the full model response to the experimental data throughout the loading process. The initial phase of the response is well captured, particularly the nonlinear evolution of the volumetric strain, which could be attributed to the initial closure of micro-cracks. However, the model does not reproduce the onset of dilatancy observed in the experiment prior to peak stress. This difference suggests the need to enrich the model with a hardening mechanism that would initiate plasticity and dissipation at an earlier stage. This potential extension, which is outlined in the perspectives section by coupling plasticity and damage, could also be beneficial to capture early hysteretic loops which can be accompanied by dilatancy \citep{Cerfontaine2018}. 

\section{Finite element simulation example} \label{sec:Goustan}

\begin{figure}[!ht]
\centering
\hspace{1.cm}
\includegraphics[width=0.6\linewidth]{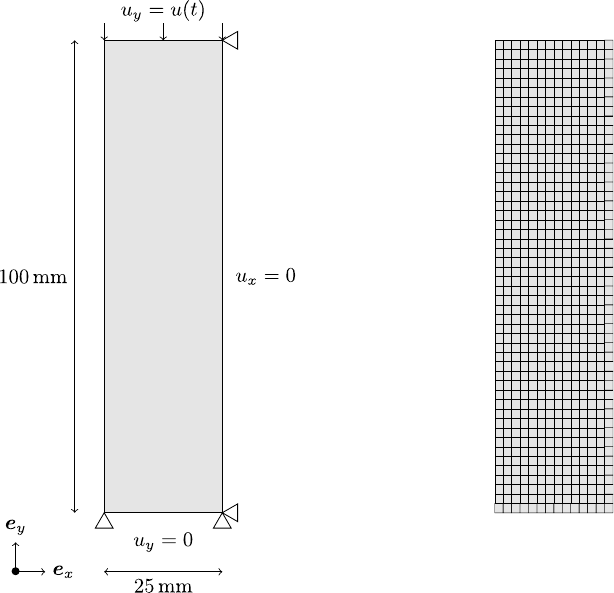}
\caption{Geometry with boundary condition and a mesh example for the compression test in plane strain conditions.}
\label{fig:geometry}
\end{figure}

To assess the numerical applicability of the proposed nonlinear elasto-plastic model, we present a 2D simulation of structural non-homogeneous responses. The problem is solved within the finite element software code\_aster \citep{aster} developed at EDF R\&D, by means of a standard Newton algorithm. We consider a compression test in plane strain condition with four structured meshes composed of quadrangular quadratic elements of size $h_c = h_0/2^n$ with $n=0,1,2,3$ and $h_0=\SI{4}{\milli\meter}$. The geometry and the boundary conditions are described in \Cref{fig:geometry}. On the top, the vertical displacement $u(t)$ is prescribed between $\SI{-2}{\milli\meter}$ and $\SI{2}{\milli\meter}$ for several loading cycles. On the bottom and on the right, the normal component of the displacement is zero. A weak element is positioned in the top-left corner by decreasing the plastic parameters $a$ and $b$ in the \Cref{tab:param_model} by $1\%$.

\clearpage

\begin{table}[ht]
\begin{center}
\begin{tabular}{||c c c c c||} 
 \hline
 $E$ & $\nu$ & $\beta_m$ & $a$ & $b$ \\
 \hline\hline
 $100$ MPa & $0.3$ & $120/E$ & $1/9$ & $E/3000$\\ 
 \hline
\end{tabular}
\end{center}
\caption{Material parameters of the nonlinear elasto-plastic model for the compression test in plane strain conditions.}
\label{tab:param_model}
\end{table}



\begin{figure}[ht]
\begin{subfigure}{0.49\textwidth}
\centering
\resizebox{!}{0.75\textwidth}{
\includegraphics{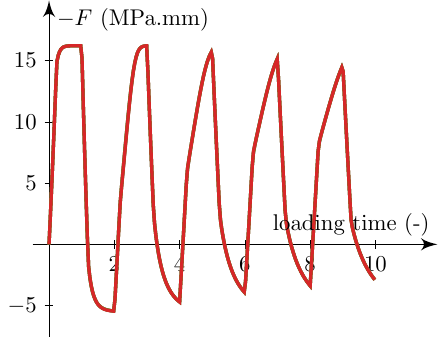}
}
\end{subfigure}
\hfill
\begin{subfigure}{0.49\textwidth}   
\centering
\resizebox{!}{0.75\textwidth}{%
\includegraphics{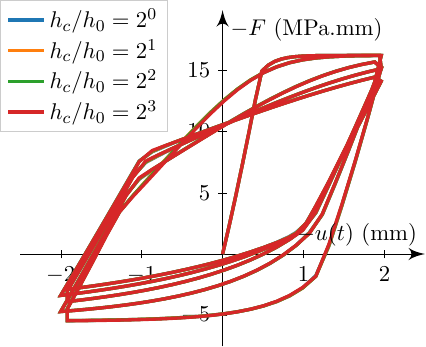}
}
\end{subfigure}
\caption{Responses of the vertical force (positive in traction) vs loading time ({\it left}) and prescribed displacement ({\it right}).}
\label{fig:response_force}
\end{figure}

\begin{figure}[ht]
\begin{subfigure}[t]{0.24\textwidth}
\centering
\includegraphics[height=7cm]{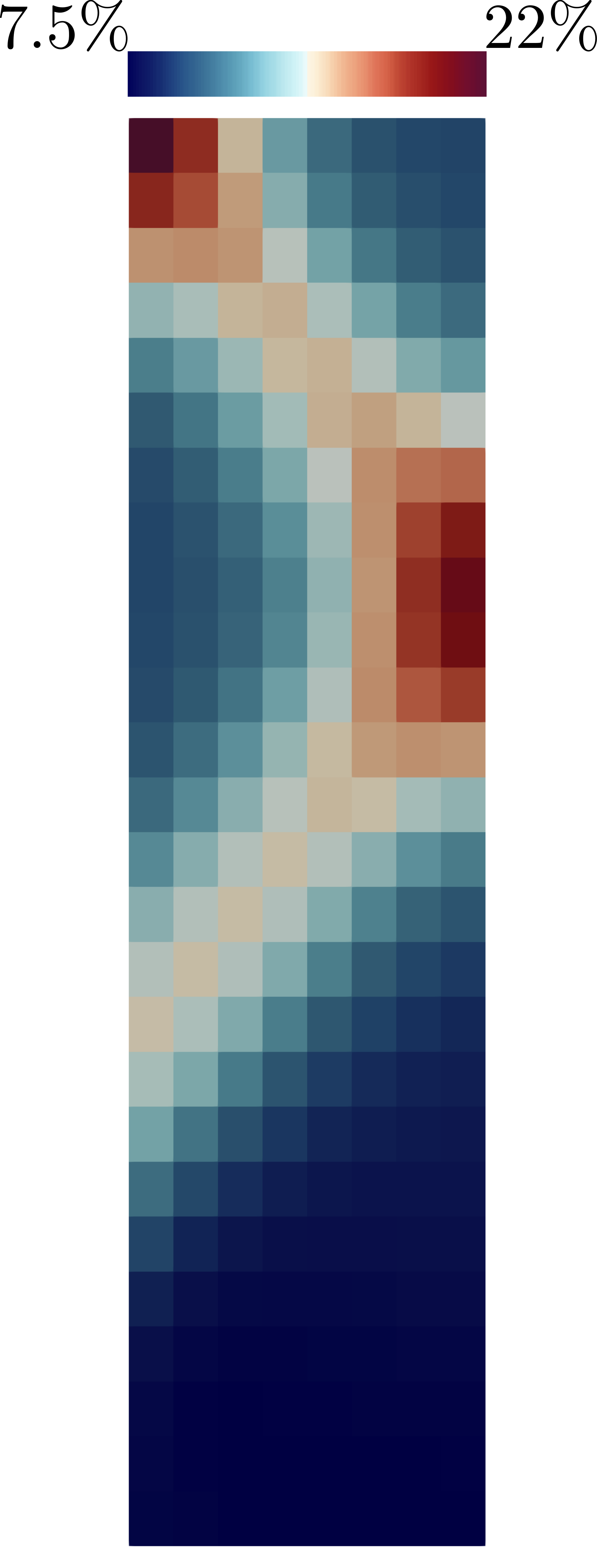}
\caption{$h_c/h_0=2^0$.}
\label{fig:iso_hc_4}
\end{subfigure}
\hfill
\begin{subfigure}[t]{0.24\textwidth}
\centering
\includegraphics[height=7cm]{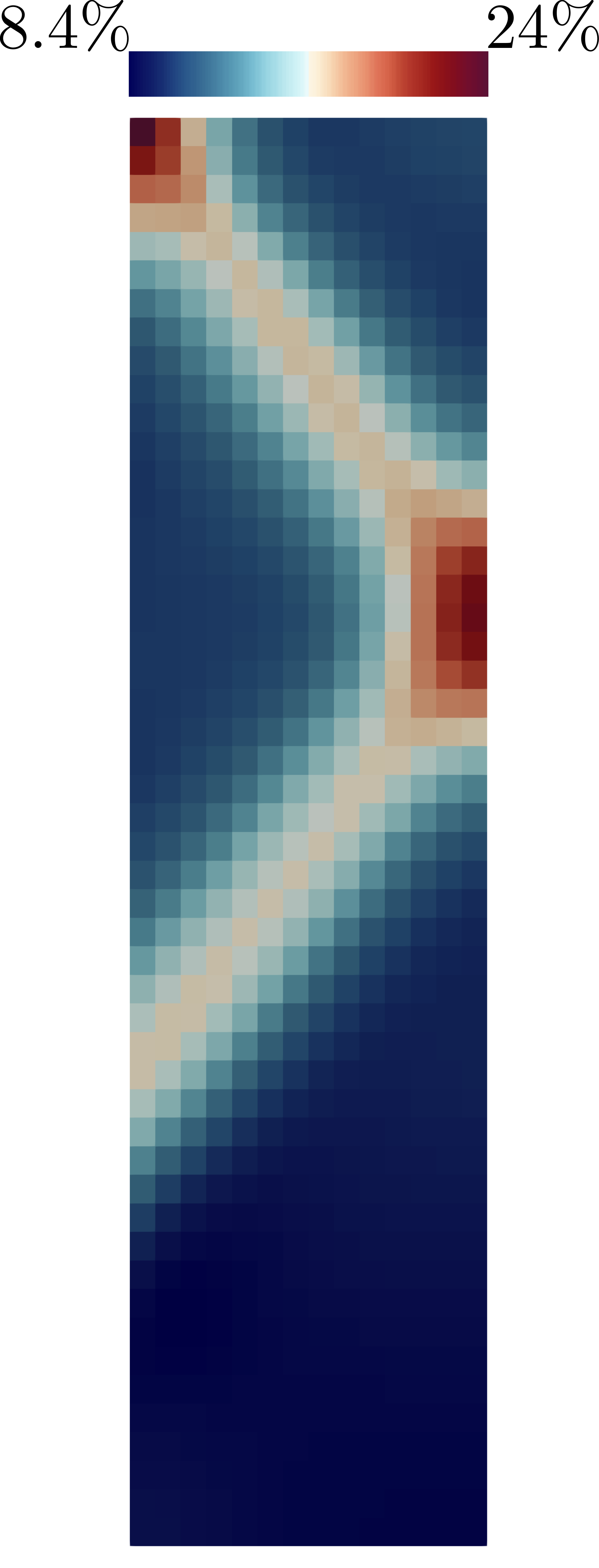}
\caption{$h_c/h_0=2^1$.}
\label{fig:iso_hc_2}
\end{subfigure}
\begin{subfigure}[t]{0.24\textwidth}
\centering
\includegraphics[height=7cm]{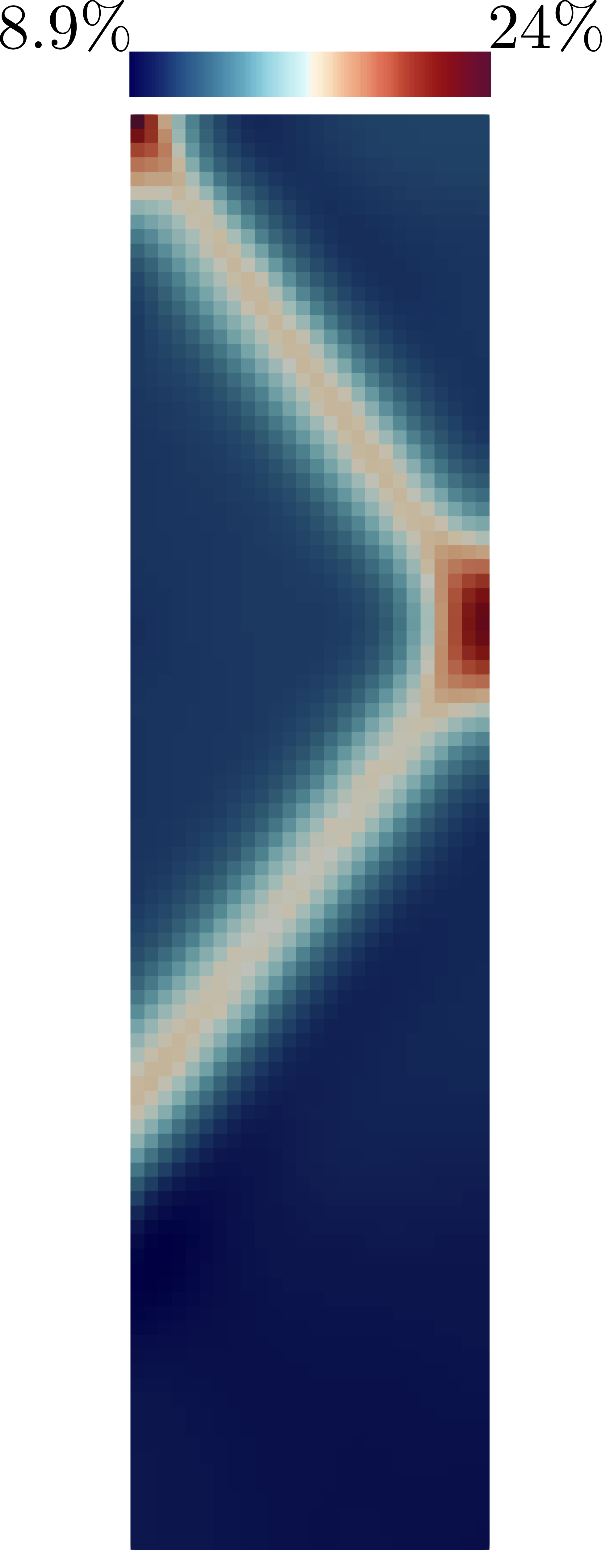}
\caption{$h_c/h_0=2^2$.}
\label{fig:iso_hc_1}
\end{subfigure}
\begin{subfigure}[t]{0.24\textwidth}
\centering
\includegraphics[height=7cm]{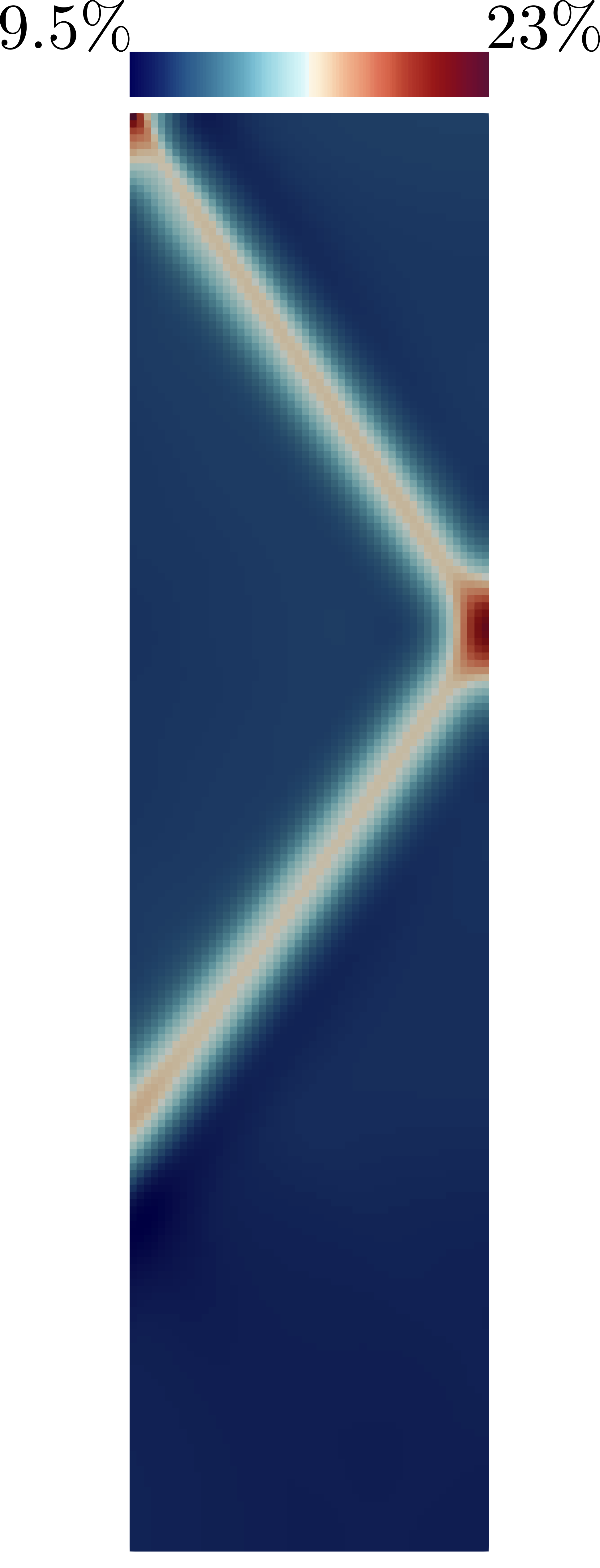}
\caption{$h_c/h_0=2^3$.}
\label{fig:iso_hc_0.5}
\end{subfigure}
\caption{Contour plots of the von Mises equivalent strain $\varepsilon_{eq}=\sqrt{2/3}\|\VEC{\varepsilon}^D\|$ at the end of the test.}
\label{fig:isovalues}
\end{figure}

\begin{figure}[!ht]
\centering
\includegraphics[height=7cm]{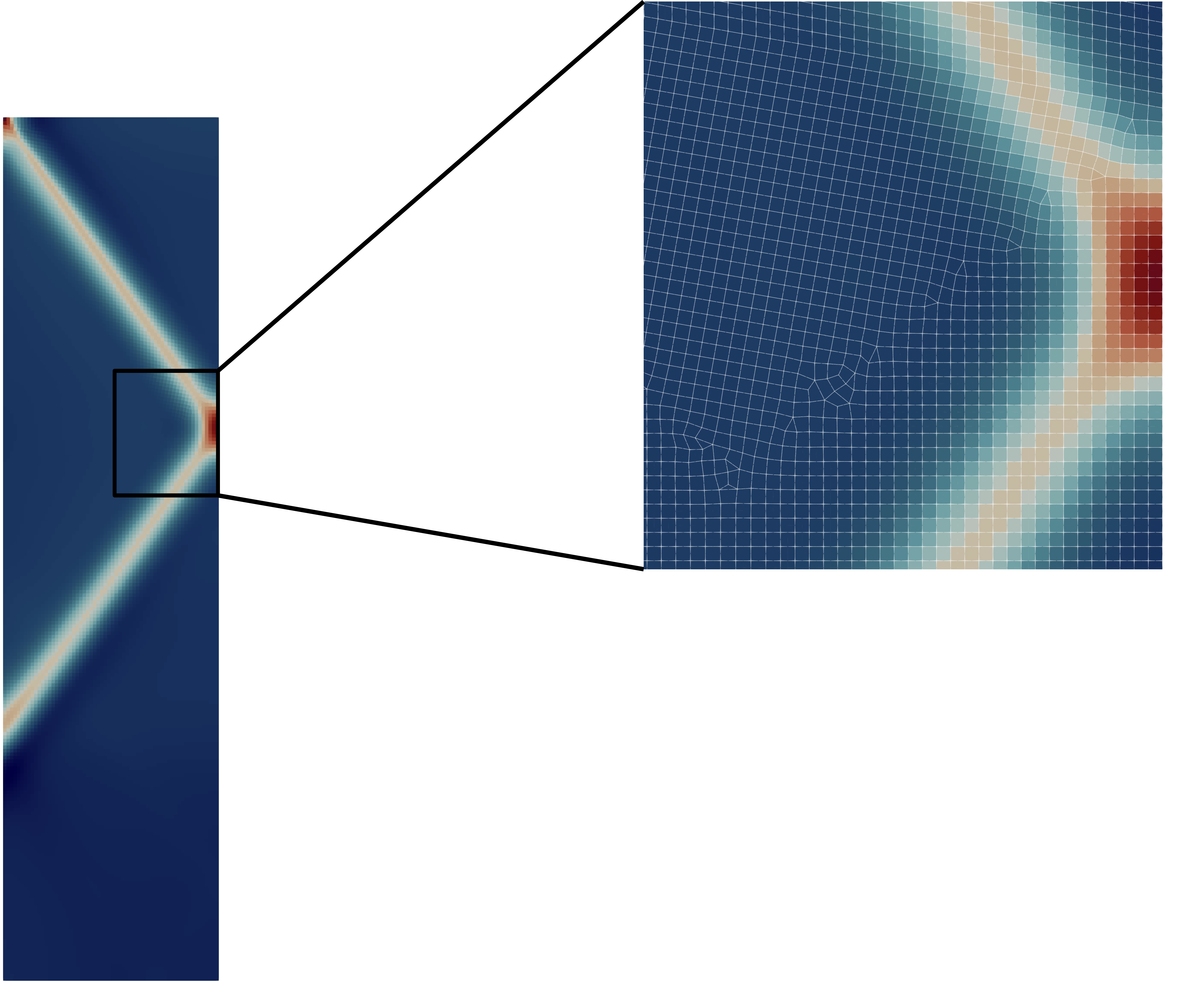}
\caption{Close–up of the shear band in case of on heterogeneously oriented mesh.}
\label{fig:orientation_mesh}
\end{figure}

\Cref{fig:response_force} shows the response of the resultant force $F$ at the top. No mesh dependency is observed. As previously presented on the material point response under a cyclic triaxial test, a progressive decay of the elastic stiffness can be observed due to positive volumetric plastic strain increments, which gradually approach a more stabilized evolution. \Cref{fig:isovalues} shows the contour plots of the von Mises equivalent strain $\varepsilon_{eq}=\sqrt{2/3}\|\VEC{\varepsilon}^D\|$ at the end of the last cycle for the four meshes considered. 
Localized deformation bands can be visualized starting from the weak element on the top-left corner of the rectangle, which lies around one element. This observation is consistent with the presence of discontinuous bifurcation modes, which are known to occur in perfect plasticity under plane strain conditions. Such behavior is typically associated with the loss of ellipticity in the governing equations \citep{Besson2010}. 
The bands are found here to have inclinations of around $53\degree-56\degree$ with the horizontal axis, which is not the inclination of the diagonal ($45\degree$) of the quadrangular elements. Finally, \Cref{fig:orientation_mesh} shows the previously identified band on a mesh with quadrangular mesh oriented by $10\degree$ in the upper part of the rectangle, demonstrating that the inclination of the band is unbiased by the mesh used.

\clearpage

\clearpage

\section{Conclusion}
In this paper, we have studied the influence of nonlinear hyperelasticity on a classical perfect plasticity constitutive behavior.  We have established the nonlinear transformation in \cref{eq:relation between stress and X new} from the plasticity thermodynamical force to the experimentally observable mechanical stresses, which respects elastic domain invariance with plastic evolution. For a particular class of hyperbolic elasticity, the hyperelastic coupling creates a link between the linear Drucker--Prager criterion in the plastic force space and a quadratic one in the stress space. The model has been constructed under the scope of the Generalized Standard Materials (GSM). Predictions of the presented model have been investigated both on the material point and on a 2D structure by using the code generation tool MFront and the finite element software code\_aster. Notably, the obtained model can exhibit dilatancy accommodation during cyclic triaxial compression tests under confining pressure, a phenomenon commonly observed in many geomaterials. This saturation of dilatancy serves as indirect evidence of the significance of nonlinear hyperelasticity. While this simplified model presented here may not capture all aspects of the complex behavior of real geomaterials,
\citep{Vermeer1984} 
it can be considered as an initial constitutive relation, serving as a fundamental brick for more subtle models, for instance, with the help of hardening mechanisms. Finally, simulations of shear band formations on highly refined meshes have proven the robust numerical behavior of the model at the structure level. The authors foresee some further investigations that the nonlinear hyperelasticity would provide, such as by analyzing the case $\beta^D>0$ in \cref{eq3:compressibility and shear moduli} to take into account nonlinear shear elasticity coupled with volumetric strain, as well as to consider previously proposed damage coupled to plasticity laws \citep{Marigo2019,Fontana2022these}. 
This extension would require furthermore the need of regularization techniques, such as gradient plasticity or gradient damage models, which can still be formulated within the GSM framework; see, for instance, \citep{Lorentz1999, Pham2011, Nguyen2021}. 
This is the topic of current research for future publications.

\section*{Acknowledgements}

The authors would like to acknowledge F. Escoffier, S. Raude (EDF R\&D Palaiseau, France), and C. Stolz (IMSIA, CNRS, France) for numerous fruitful discussions. We thank the CIH team of EDF Hydro (Chambéry, France) for their support and technical advice. 

\appendix

\section{Hoek--Brown type quadratic yield criteria} \label{sec:HB_history}

We summarize here the brief history of the introduction of quadratic yield surfaces, what we call ``Hoek--Brown type'' yield surfaces.
Due to discrepancies between the experimental results and those predicted by Mohr--Coulomb criterion, in the early eighties scientific community came progressively to the idea that a suitable shape of yield surfaces for rock materials has to be a curved, apex-pointed cone with eventually curved cross sections in the octahedral planes. The Hoek–Brown strength criterion \citep{Hoek1980} was specifically developed at that time for characterizing the mechanical behavior of rock materials and rock masses. Its consequent widespread adoption in engineering practice was largely due to two key advantages: (1) it accurately captured the quadratic relationship between shear strength and hydrostatic compression, and (2) it relied on a limited set of input parameters that can be conveniently obtained from standard compression tests and sample mineralogical examination. We focus here mainly on a single aspect of the Hoek–Brown criterion extension that is most relevant to our study: its parabolic reformulation in terms of rotational invariants \citep{Pan1988}. A comprehensive discussion of its theoretical foundations, the various forms of 3D extensions \citep{Zhang2013}, experimental validation \citep{Li2021}, full parameter calibration, and detailed practical applications \citep{Hoek2019} can be found in the literature, but lies beyond the scope of the present paper. \\

Back in 1924, while studying the mechanical behavior of glasses, Griffith \citep{Griffith1924} was the first to derive from theoretical considerations a quadratic multi-axial criterion for fracture
\begin{equation}
(\stress_1 - \stress_3)^2 \sim \stress_1 + \stress_3,
\end{equation}
where $\stress_1$ and $\stress_3$ are the major and minor principal stresses, respectively. Forty years later, Fairhurst \citep{Fairhurst1964} attempted to empirically extend the work of Griffith to the domain of high compression suitable for rock behavior analysis. Finally, in 1980, Hoek and Brown \citep{Hoek1980} obtained a criterion shape that convinced many generations of geomaterial scientists. The original Hoek--Brown criterion is still widely used in rock mechanics, and, for intact rocks, it can be written as
\begin{equation}
\stress_1 = \stress_3 + C_0\sqrt{m_i\frac{\stress_3}C_0+1},
\end{equation}
where $C_0$ is the uniaxial compressive strength and $m_i$ is a material constant for the intact rock. For more details about these constants, we refer to \citep{Hoek2019}, where a generalized version of the criterion involving the geological strength index (GSI) is also proposed and analyzed. Furthermore, the evolution of the Hoek--Brown criterion in the literature can be found, summarized, in \citep{Hoek2007}. \\

Since many papers have exhibited the strong influence of the intermediate principal stress $\stress_2$ (see e.g. \citep{Eberhardt2012} and the references therein),
multiple three-dimensional extensions of Hoek--Brown criterion have been developed, either based directly on principal stresses  (see e.g. \citep{Zhang2013, Li2021} and the references therein) or formulated in terms of rotational invariants \citep{Pan1988,Priest2012}. These formulations were further employed as building blocks for more advanced evolving post-peak constitutive relations as in \citep{Pan1988,Raude2016}. The work of X. D. Pan and J. Hudson was not solely motivated by the 3D extension of the model; the authors also emphasized the importance of capturing post-failure behavior and the practical advantage of using a model with a minimal number of parameters for excavation design. Its first generalized form, written with the help of rotational invariants, was proposed in \citep{Pan1988} and preserved Hoek--Brown type parametrization, highlighting advantageous, easier finite element implementation:
\begin{equation}
\frac{3}{2C_0} \left\lVert\stressb^D\right\rVert^2 + \frac{\sqrt{3}m_i}{2\sqrt{2}} \left\lVert\stressb^D\right\rVert + m_i\stress_m - C_0 \leq 0.
\end{equation}
Many other, derived from Hoek--Brown, criteria exist, but most of them keep the initial nearly parabolic dependence and can be approximated by the general form invariant-based expression, that we are investigating in this paper:
\begin{equation}\label{eq:PH criterion}
f_{\stressb} = \left\lVert\stressb^D\right\rVert^2 + A \left\lVert\stressb^D\right\rVert + B \stress_m - C \leq 0.
\end{equation}
Although the Hoek--Brown criterion was proposed nearly half a century ago, its extension remains an active area of research, continuing to inspire numerous studies and being supported by various experimental findings. We hope that our contribution (compare, for instance, \cref{eq:PH criterion} and \cref{eq:criterion in stress space} in the main text) offers a new way capable of explaining this experimentally observed curved yield surface within a novel approach.

\section{Numerical integration procedure of the model}\label{sec:numerical_integration}

We use an implicit Euler method 
to integrate the equations of the proposed nonlinear elasto-plastic model. The linear Drucker--Prager type flow rule advantageously enables the integration scheme to produce a fully analytical expression. \\

Given the total strain $\strainb$ at the current time increment $t_{n+1}$ and the plastic strain $\plb_n$ at the previous one $t_{n}$, the main numerical integration goal is to determine the plastic strain increment $\Delta\plb$, the stress $\stressb$, and the consistent tangent operator ${d\stressb}/{d\strainb}$. To this end, we employ a classical return-mapping scheme. The procedure begins by assuming that the strain increment is purely elastic. The plastic force is thus
\begin{equation}\label{eq:elastic_trial}
    \plforb = \plforb^{el} := \kappa(\strainb)\tr(\strainb-\plb_n)\idtwo + 2\mu(\strainb^D-\plb_n^D).
\end{equation}

We omit the subscript $n+1$ on all quantities evaluated at the current time increment. If the yield function satisfies $f_{\plforb}(\plforb^{el})\leq 0$, the strain increment is indeed elastic, and no plastic evolution occurs. However, if $f_{\plforb}(\plforb^{el})>0$, plastic correction is required. In this case, two scenarios must be carefully distinguished: plasticity may occur either on the smooth portion of the Drucker–Prager yield surface, or at its apex.\footnote{For the determination of the case to consider during the plastic correction, either on the smooth portion of the Drucker--Prager yield surface or at its apex, we use the decision criterion established in \citep{Sysala_2016}.}

\subsubsection*{Increment of the plastic strain on the smooth portion}

On the smooth portion of the yield surface, the deviatoric part of the plastic force does not vanish. By taking into account the state equation \eqref{eq:X expression new}, the flow rule \eqref{eq:flow rule X}, and the elastic trial \eqref{eq:elastic_trial}, the plastic force is expressed
\begin{equation}
    \plforb = \plforb^{el} - 2\mu\Delta\plb^D -\kappa(\strainb)\tr(\Delta\plb)\idtwo = \plforb^{el} - \Delta\lambda\left(2\mu\frac{1}{\sqrt{6}}\frac{\plforb^D}{\| \plforb^D\|}+\kappa(\strainb)a\idtwo\right),
\end{equation}
where, we note by $\Delta \lambda$ the increment of the plastic multiplier between $t_n$ and $t_{n+1}$. By taking the deviatoric part of this relationship, we have $\VEC{n}^D := \plforb^D/(\sqrt{6}\|\plforb^D\|) = \plforb^{D,el}/(\sqrt{6}\|\plforb^{D,el}\|)$, i.e., the direction of the deviatoric plastic strain is determined by the elastic trial. Then, by expressing $f_{\plforb}(\plforb) = 0$ from \cref{eq:Drucker-Prager criterion X}, we obtain the increments $\Delta\lambda$ and $\Delta\plb$ as
\begin{equation}
    \Delta\lambda = \frac{f_{\plforb}(\plforb^{el})}{\mu/3 + a^2\kappa(\strainb)},\quad
    \Delta\plb = \Delta\lambda\left(\VEC{n}^D+\frac{a}{3}\idtwo\right).
\end{equation}

\subsubsection*{Increment of the plastic strain at the apex}

At the apex of the yield surface, the deviatoric part of the plastic force is zero, i.e., $\plforb^{D} = \plforb^{D,el} - 2\mu\Delta\plb^D = \VEC{0}$, which directly determines the increment of the deviatoric part of the plastic strain. Moreover, $\plform = \plform^{el} - \kappa(\strainb)\tr(\Delta\plb)$. Subsequently, by enforcing $f_{\plforb}(\plforb) = 0$, that reduces to $a\plform-b=0$, we obtain
\begin{equation}
    \Delta\plb = \frac{\plforb^{D,el}}{2\mu} + \frac{a \plform^{el} - b}{3a\kappa(\strainb)}\idtwo.
\end{equation}

\subsubsection*{Expression of the stress and of the consistent tangent operator}

From \cref{eq:stress expression new}, the stress can be written
\begin{equation}
    \stressb = \hat {\stressb}(\strainb,\plb) := \kappa(\strainb)\tr(\strainb-\plb)\left[1-\beta_m\kappa(\strainb)\tr(\strainb-\plb)\right]\idtwo + 2\mu(\strainb^D-\plb^D),
\end{equation}
with $\plb=\plb_n+\Delta\plb$. The consistent tangent operator can therefore be computed with the help of the chain rule
\begin{equation}
    \frac{d\stressb}{d\strainb} = \frac{\partial \hat {\stressb}}{\partial \strainb} + \frac{\partial \hat {\stressb}}{\partial \plb}:\frac{\partial \plb}{\partial \strainb}.
\end{equation}

The resulting expressions, written with help of shorthand notations of the fourth-order spherical and deviatoric projector tensors $\MAT{J}:=\idtwo\otimes\idtwo/3$ and $\MAT{K}:=\MAT{I} - \MAT{J} $, give:
\begin{itemize}
    \item Elastic response:
    \begin{equation}
        \frac{d\stressb}{d\strainb} =
        3\kappa(\strainb)\omega(\strainb,\plb)^2\MAT{J} + 2\mu\MAT{K}, \quad\text{with}\quad \omega(\strainb,\plb) = \frac{1+2\beta_m\kappa_i\tr\plb}{1+2\beta_m\kappa_i\tr\strainb}.
    \end{equation}
    \item Plastic response--smooth portion of the yield surface:
    \begin{align}
        \frac{d\stressb}{d\strainb} =&\  
        3\kappa(\strainb)\omega(\strainb,\plb)^2\MAT{J} + 2\mu\MAT{K}\nonumber \\
        &- \frac{\left[a\kappa(\strainb)\omega(\strainb,\plb)\idtwo+2\mu \VEC{n}^D\right]\otimes\left[a\kappa(\strainb)\omega(\strainb,\plb)\idtwo+2\mu \VEC{n}^D\right]}{\mu/3 + a^2\kappa(\strainb)}
        \nonumber\\
        &-\frac{\Delta\lambda}{\sqrt{6}}\frac{4\mu^2}{\|\plforb^{D,el}\|}\left(\MAT{K}-6\VEC{n}^{D}\otimes\VEC{n}^{D}\right).
    \end{align}
    \item Plastic response--apex of the yield surface:
    \begin{equation}
    \frac{d\stressb}{d\strainb} = 0.
\end{equation}
\end{itemize}

In all cases, the consistent tangent operator is symmetric, as expected from the Generalized Standard Materials (GSM) framework. For the elastic response, the operator is positive-definite provided that $\kappa(\strainb)>0$. At the apex, the consistent tangent operator reduces to the zero tensor, which is a classical result for the Drucker--Prager criterion without hardening \citep{Sysala_2016}. For the plastic response on the smooth portion of the yield surface, we have, for any symmetric second-order tensor $\VEC{\eta}$,
\begin{align}
    \VEC{\eta}:\frac{d\stressb}{d\strainb}:\VEC{\eta} =&\  \kappa(\strainb)\omega(\strainb,\plb)^2(\tr\VEC{\eta})^2 + 2\mu\|\VEC{\eta}^D\|^2 - \frac{\left(a\kappa(\strainb)\omega(\strainb,\plb)\tr\VEC{\eta} + 2\mu\VEC{\eta}^D:\VEC{n}^D\right)^2}{\mu/3+a^2\kappa(\strainb)} - \\
    &- \frac{\Delta\lambda}{\sqrt{6}}\frac{4\mu^2}{\|\plforb^{D,el}\|}\left[\|\VEC{\eta}^D\|^2-6(\VEC{\eta}^D:\VEC{n}^D)^2\right]. \nonumber
\end{align}

Using the inequality $\Delta\lambda < \sqrt{6}\|\plforb^{D,el}\|/(2\mu)$ because $\|\plforb^{D}\|>0$ on the smooth portion of the yield surface, and noting that the final bracketed term is non-negative, we conclude
\begin{equation}
    \VEC{\eta}:\frac{d\stressb}{d\strainb}:\VEC{\eta} \geq 
    \frac{\mu\kappa(\strainb)/3}{\mu/3+a^2\kappa(\strainb)}\left(\omega(\strainb,\plb)\tr\VEC{\eta}-6a\VEC{\eta}^D:\VEC{n}^D\right)^2 \geq 0,
\end{equation}
which proves the semi-definite-positiveness of the consistent tangent operator.


\clearpage

\bibliographystyle{elsarticle-harv} 
\bibliography{refs}

\end{document}